\documentclass[preprint,12pt]{elsarticle}
\usepackage{threeparttablex}
\usepackage{booktabs}
\usepackage{amsmath}
\usepackage{lipsum}

\usepackage{amssymb}

\begin{document}

\begin{frontmatter}



\title{On the analysis of signal peaks in pulse-height spectra}


\author{Cade R. Rodgers}
\address{Department of Physics \& Astronomy, University of North Carolina at Chapel Hill, NC 27599-3255, USA}

\author{Christian Iliadis}
\address{Department of Physics \& Astronomy, University of North Carolina at Chapel Hill, NC 27599-3255, USA}
\address{Triangle Universities Nuclear Laboratory (TUNL), Durham, North Carolina 27708, USA}

\begin{abstract}
The estimation of the signal location and intensity of a peak in a pulse height spectrum is important for x-ray and $\gamma$-ray spectroscopy, charged-particle spectrometry, liquid chromatography, and many other subfields. However, both the ``centroid'' and ``signal intensity'' of a peak in a pulse-height spectrum are ill-defined quantities and different methods of analysis will yield different numerical results. Here, we apply three methods of analysis. Method A is based on simple count summation and is likely the technique most frequently applied in practice. The analysis is straightforward and fast, and does not involve any statistical modeling. We find that it provides reliable results only for high signal-to-noise data, but has severe limitations in all other cases. Method B employs a Bayesian model to extract signal counts and centroid from the measured total and background counts. The resulting values are derived from the respective posteriors and, therefore, have a rigorous statistical meaning. The method makes no assumptions about the peak shape. It yields reliable and relatively small centroid uncertainties. However, it provides relatively large signal count uncertainties. Method C makes a strong assumption regarding the peak shape by fitting a Gaussian function to the data. The fit is based again on a Bayesian model. Although Method C requires careful consideration of the Gaussian width (usually given by the detector resolution) used in the fitting, it provides reliable values and relatively small uncertainties both for the signal counts and the centroid.
\end{abstract}




\end{frontmatter}


\section{Introduction} \label{sec:intro}
The reliable estimation of the signal location and intensity of a peak in a pulse height spectrum is of paramount importance for many analytical techniques, including x-ray and $\gamma$-ray spectroscopy, charged-particle spectrometry, and liquid chromatography. The goal of measurements is to accumulate ``good statistics'' on a peak of interest, so that the peak parameters can be extracted. But both the ``centroid'' and ``signal intensity'' of a peak are ill-defined quantities and one can reasonably expect that different methods of analysis will yield different numerical results. These quantities are then defined by the technique used to determine them, and the quality of a given method will depend on how reproducible the results are \cite{deb88}. The simplest techniques estimate the peak parameters based on the data alone, without involving any other assumptions on statistical modeling. More involved techniques resort to suitable statistical models. In practice, the problem of estimating the signal centroid and intensity of a peak is lessened because experimental energy and efficiency calibrations are presumably determined with the same analysis method, and, therefore, any systematic bias is reduced in the energy and efficiency corrected quantities.

The problem is exacerbated when the peak is superimposed on a significant background. In this case, the total number of counts recorded by the detector is not an accurate representation of the signal intensity, because various sources of background also contribute to the total number of counts in the region of the peak whose net intensity we seek to estimate. Efficiencies are usually determined from calibration peaks that rise far above the background. If the peak of interest is superimposed on a relatively large background, the ratio of these two peak intensities will be subject to the systematic bias referred to above. 

Previously, data have been analyzed using the method of least squares to fit a parameterized peak shape function plus background to a region about the peak. Phillips \cite{PHILLIPS1978449} pointed out that the application of such techniques to data of low statistics can result in fits that are significantly biased toward too low or too high a peak area and also yield unreliable peak positions. He suggested modifications to the least-squares method of traditional statistics to reduce the bias. He also found that for isolated peaks a simple ``hand analysis'' provided superior results compared to more sophisticated methods. Instead of a least-squares analysis, Awaya \cite{AWAYA1979317} employed a maximum likelihood method using Poissonian probability distributions, which was found to give reliable results. A similar approach, but with special emphasis on small signals, was presented by Hannam and Thompson \cite{HANNAM1999239}. 

In the present work, we will revisit the problem of peak analysis in the presence of a background. Previous work was mainly focused on the analysis of mean values for the signal peak centroid and intensity, whereas we are equally interested in analyzing reliable {\it uncertainties} for these quantities. To this end, we employ three methods. Method A is based on count summation and is similar to the ``hand analysis'' of Ref.~\cite{PHILLIPS1978449}. Method B involves a bin-by-bin analysis using a Bayesian model. Method C has similarities to the model discussed in Ref.~\cite{AWAYA1979317}, but is again based on a Bayesian model. We are not concerned here with upper bound\footnote{For the important distinction between ``upper bounds'' and ``upper limits'', see Ref.~\cite{Kashyap_2010}. The former describe the result of a measurement and depend on the number of counts in both the peak and background regions. The latter are specific to the measurement technique, involve the number of background (but not signal) counts, and can be estimated before knowing the experimental outcome.} estimates of signal counts (see, e.g., Ref.~\cite{Zhu:2007gf}), although some of the methods discussed here could be useful for this purpose as well. Instead, we focus on the analysis of data when a peak can be discerned in the presence of background. We will assume that the data are subject to statistical uncertainties only. In particular, we will disregard the impact of any systematic effects on the data, such as wrong instrument calibrations, faulty adjustments of the experimental setup, or incorrect judgements when making observations.

In this work, we will analyze computer-generated data only. In real experiments, the true parent distribution from which the data are extracted is unknown. This vastly complicates the testing of different analysis methods. However, for computer-generated data the parameters of the parent distribution are precisely known and we can thoroughly test different analysis techniques by checking if they recover the parameters originally used in the artificial data generation.

In Section~\ref{sec:test} we discuss our method of generating artificial data. Different methods of peak analysis are described in Section~\ref{sec:method}. Our results are presented in Section~\ref{sec:results}. A concluding summary is given in Section~\ref{sec:summary}.

\section{Test data} \label{sec:test}
Before we can test different methods of analyzing peaks in pulse-height spectra, we need to generate artificial data with a precisely known peak centroid and intensity. The usefulness of any analysis method will depend on its ability to reproduce these values from the input of the artificial data only. We will generate the data assuming a Gaussian peak shape and a constant background level. The extension to other peak shapes and background assumptions (e.g., exponential or step functions caused by multi-Compton events and other physical processes) is straightforward but beyond the scope of this work.

In a real nuclear counting measurement, the number of decaying nuclear levels (``trials'') is high and either the probability of decay per nucleus or the efficiency of detection per event (``success'') is low. In such cases, counting statistics is given by a Poissonian, which relates the true number of total counts (expectation value) to the observed number of total counts. For a hypothetical detector of infinite energy resolution, the observed number of total counts would all appear in a single bin. In a real detector, random noise is added to each of these counts, where the noise is sampled from a Gaussian distribution with a standard deviation that reflects the detector resolution. The above sequence only refers to the signal. In each channel, a background contribution, which is separate of the signal, is then added. Notice that the above sequence involves the independent sampling from two distributions: a Poissonian, which gives rise to the difference between total true and observed counts, and a Gaussian, which distributes the counts along the channel direction.

We generated artificial data by following these steps: 

(i) The user chooses ``true'' values, which will be unknown to the analysis method, for the centroid and width of a Gaussian, and the total number of signal counts (area of Gaussian). 

(ii) Random samples are drawn from the specified Gaussian distribution and are binned to produce a signal histogram. 

(iii) A background histogram is produced, with the same number of channels and a constant (``true'') number of background counts in each bin. 

(iv) For each bin in the signal histogram, the number of counts is adopted as the mean value to draw random samples from a Poissonian distribution to generate a new signal histogram. We refer to the peak centroid and total area of this set as the ``actual'' signal values. A similar procedure is applied to the background histogram. 

(v) These two histograms are added bin-by-bin to generate the final histogram containing the artificial (``observed'') data to be analyzed.

The artificial data\footnote{Notice that we reversed the sampling order in the generation of artificial data (first from a Gaussian, then a Poissonian) compared to the accumulation of real data (first from a Poissonian, then a Gaussian). The order is inconsequential because both processes are independent, but the first choice is computationally simpler.} are suitable for testing the reproducibility of an analysis method because all steps are precisely controlled. No approximation is introduced by adding the two histograms in the last step because the sum of two independent Poissonian random variables is also a Poissonian random variable. Real life data correspond only to the ``observed'' values, as defined above, and both their ``true'' and ``actual'' values are only known to mother nature. 

\begin{figure}[htbp]
\includegraphics[width=0.65\columnwidth]{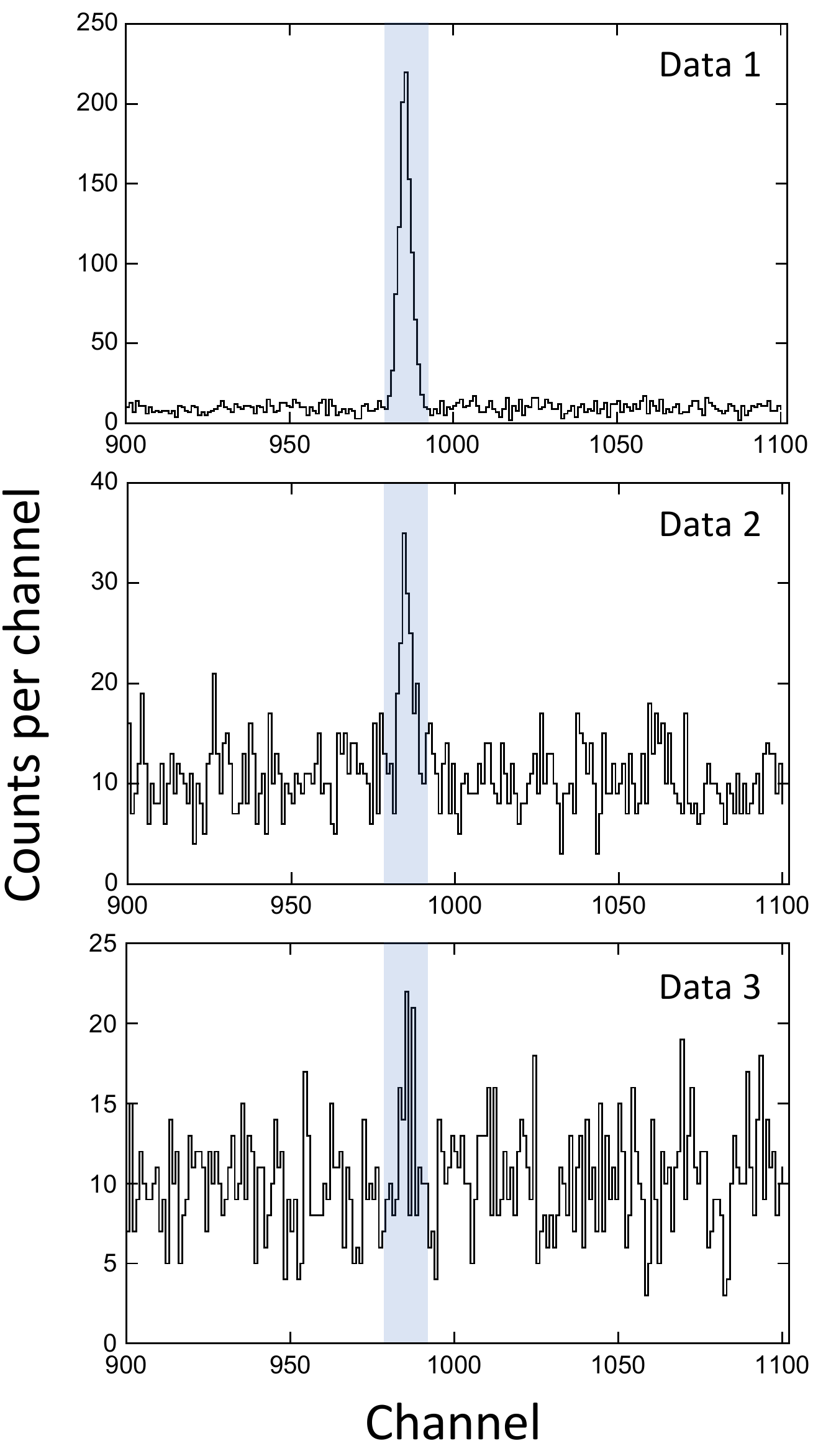}
\caption{\label{fig:data} Test data generated according to the procedure outlined in Section~\ref{sec:test}. They reflect high (top), moderate (middle), and low (bottom) signal-to-background ratios. In all panels, the true peak centroid is channel $985.0$, the true full-width-half-maximum (FWHM) is 4.7 channels, and the true background is 10 counts per channel. The data shown in the top, middle, and bottom panels were generated assuming $970$, $105$, and $35$ true signal counts, respectively. The peak region, shaded blue, is 14 channels wide, corresponding to three times the FWHM.
}
\end{figure}

Figure~\ref{fig:data} depicts three data sets with varying signal-to-noise ratios that we generated for testing. They roughly represent high, moderate and low signal-to-background ratios. All data sets shown were obtained with a true peak centroid of channel $985.0$, a Gaussian standard deviation of 2.0 channels, corresponding to a full-width-half-maximum (FWHM) of 4.7 channels, and a true background of 10 counts per channel. The data shown in the top, middle, and bottom panels (Data 1, 2, and 3, respectively) contain $970$, $105$, and $35$ true signal counts, respectively. Our task is to estimate from the ``observed'' data the signal peak centroids and intensities, together with their associated uncertainties, and determine if our results agree with the values used to generate the data.

\section{Methods} \label{sec:method}
We will discuss three different methods of peak analysis and compare the results. Method~A relies on simple count summation and analyzes the data without any assumptions about statistical models. Method~B is based on a Bayesian model without any assumptions about the peak shape. Method~C performs a Gaussian fit to the data, again employing a Bayesian model. 

For all three methods, we will assume that the signal only occurs in a user-defined peak region. We also need a model for the background because it is not separately known from the number of counts in the peak region alone. It can be approximated by choosing signal-free background regions to the left and right of the peak. We will assume for all three methods a linear background over the peak region. The extension to more complicated background assumptions is straightforward. 

The average background, $B_i$, in channel $x_i$ can be linearly interpolated by using the means of the two background region locations, ${x}_{B1}$ and ${x}_{B2}$, and the corresponding mean background counts per channel, ${B}_{B1}$ and ${B}_{B2}$: 
\begin{equation} \label{eq:bg}
B_i = \left( \frac{B_{B2} - B_{B1}} {x_{B2} - x_{B1}} \right) (x_i - x_{B1}) + B_{B1}
\end{equation}
The above expression will be used for Methods A and B to estimate the background in the peak region. For Method C, the slope and intercept of the linear background approximation are fitting parameters.

{\it Comment 1:} The background in the peak region is described by a straight line. For Methods A and B, it is found from Equation~(\ref{eq:bg}), which disregards statistical fluctuations in the background. 

\subsection{Method A: Count summation} \label{sec:method_A}
We will start with a simple method based on count summation, which is frequently applied in this or similar form. The total number of counts in the peak region, which is comprised of $n$ channels, is
\begin{equation}
\label{eq:total}
T = \sum_i^n T_i
\end{equation}
with $T_i$ the number of total counts in channel $x_i$. The estimated number of background counts from Equation~(\ref{eq:bg}) in the peak region is
\begin{equation}
\label{eq:bkg}
B = \sum_i^n B_i
\end{equation}
The number of signal counts in the peak region is then
\begin{equation}
\label{eq:netcounts}
S = \sum_i^n S_i = T - B
\end{equation}
Assuming that counting statistics obeys a Poissonian distribution, with the standard deviation equal to the square root of the mean value, and that the number of counts in the peak and background regions are uncorrelated, the uncertainty in the signal counts is 
\begin{equation}
\label{eq:netcountserr}
\sigma_S = \sqrt{\sigma_T^2 + \sigma_B^2} = \sqrt{T+B}
\end{equation}
where $\sigma_T$ and $\sigma_B$ denote the standard deviation in the total and background counts, respectively. 

If the number of signal counts is deemed statistically significant, the signal centroid can be estimated using the signal counts in each channel, $S_i$ $=$ $T_i$ $-$ $B_i$. We define the signal centroid as the sample mean:
\begin{equation}
\label{eq:cent}
\bar{x} = \frac{1}{S} \sum_{i}^{n} S_i x_i
\end{equation}
In other words, as counts accumulate in the peak, we can determine an average location by weighting the channel values, $x_i$, by the number of signal counts contained in the channel, $S_i$. The amount of internal fluctuation in the data is quantified by the sample standard deviation:
\begin{equation}
\label{eq:stand}
s_x^2 = \frac{1}{S-1} \sum_{i}^{n} S_i (x_i-\bar{x})^2
\end{equation}
The uncertainty in the peak centroid is estimated from the standard error of the mean:
\begin{equation}
\label{eq:centerr}
\sigma_{\bar{x}}=\frac{s_x}{\sqrt{S}}
\end{equation}
Equations~(\ref{eq:netcounts}), (\ref{eq:netcountserr}), (\ref{eq:cent}), and (\ref{eq:centerr}) are estimates for the actual number of signal counts and the peak centroid. The expressions are simple to use and do mostly not depend on assumptions about specific statistical parent distributions or the shape of the peak. 

{\it Comment 2:} In the limit of small signal counts, the uncertainty, $\sigma_S$, will become comparable to, or even exceed, the best estimate, $S$, and the result will become meaningless in a statistical sense. Also, in this case the determination of a centroid according to Equations~(\ref{eq:cent}) and (\ref{eq:centerr}) has a questionable meaning.

{\it Comment 3:} Equations~(\ref{eq:cent}), (\ref{eq:stand}), and (\ref{eq:centerr}) assume that the quantity $S_i$ has no uncertainty; however, for each channel, $S_i$ is not measured directly, but is derived from the total counts and background counts, each of which are distributed according to a Poissonian. 

{\it Comment 4:} The subtraction procedure, $S_i$ $=$ $T_i$ $-$ $B_i$, implied by Equation~(\ref{eq:netcounts}), may give rise to negative values of $S_i$. This is problematic because these negative values will introduce a bias by artificially shifting the centroid to a smaller or larger value, according to Equation~(\ref{eq:cent}), depending on where these negative-signal-count channels occur in the peak region. Furthermore, negative signal counts could give rise to a negative value of $s_x^2$ in Equation~(\ref{eq:stand}), which would prohibit the estimation of a centroid uncertainty according to Equation~(\ref{eq:centerr}). Therefore, we will arbitrarily replace any negative signal counts by zero counts for the centroid calculation only. 

{\it Comment 5:} The reproducibility of the results is questionable, i.e., changing the regions of the peak and the backgrounds to the left and right of the peak may yield values outside the uncertainties defined by Equations~(\ref{eq:netcountserr}) and (\ref{eq:centerr}). 


\subsection{Method B: Bayesian model} \label{sec:bayes}
While Method A is well-suited for high signal-to-noise ratios, Comments 2 $-$ 5 serve as  limiting factors as the number of signal counts decreases. To obtain a more reliable estimate for the observed signal counts and the peak centroid, we will adopt a Bayesian model. Bayes' theorem is given by 
\begin{equation}
\label{eq:bayes}
p(\theta | D) = \frac{p(D|\theta)p(\theta)}{\int{p(D|\theta)p(\theta)d\theta}}
\end{equation}
where $D$ and $\theta$ $=$ $(\theta_0, \theta_1,...,\theta_n)$ denote the data being analyzed and the vector of model parameters, respectively. All of the functions $p$ in Equation~(\ref{eq:bayes}) represent distinct probability densities: $p(D|\theta)$ is the ``likelihood,'' i.e., the conditional probability that the data, $D$, are collected assuming given values for the model parameters, $\theta$; the ``prior,'' $p(\theta)$, is the joint unconditional probability for a given set of model parameters before seeing the data. The product of likelihood and prior defines the quantity of main interest here, called ``posterior,'' $p(\theta|D)$, which is the joint conditional probability for a set of model parameters given the data. The denominator in Equation~(\ref{eq:bayes}), called ``evidence,'' serves as a normalization factor and is not important in the present context. For an introduction to Bayesian models, see, e.g., Ref.~\cite{2017bmad}.

To obtain the (``marginalized'') posterior probability density for a single parameter, say, $\theta_0$, we integrate over all other parameters,
\begin{equation}
\label{eq:bayes2}
p(\theta_0 | D) = \int_{\theta_1,\theta_2,...,\theta_n} p(\theta | D) d\theta_1 d\theta_2 ... d\theta_n
\end{equation}
The quantity $p(\theta_0 | D)$ summarizes all of our knowledge about $\theta_0$, given both the prior information and the new data, $D$. Meaningful parameter values are obtained from this expression by defining ``credible'' intervals, $[\theta_a,\theta_b]$, according to
\begin{equation}
\label{eq:bayes3}
\int_{\theta_a}^{\theta_b} p(\theta_0 | D) d\theta_0 = \beta
\end{equation}
where $\beta$ is a user-defined probability. We define parameter values and their uncertainties by using 16, 50, and 84 percentiles, corresponding to a credibility level\footnote{In nuclear physics, uncertainties are usually presented as one ``standard uncertainty'' \cite{jcgm:2008:EMDG,Audi_2017}. For example, the $\gamma$ ray energies and emission probabilities presented in the Evaluated Nuclear Structure Data File (ENSDF) \cite{tuli} refer to one standard uncertainty. If the underlying probability density is a Gaussian, then the standard uncertainty has a 68\% credibility level. This is the reason why we also adopt this value in our Bayesian model.} of $\beta$ $=$ $68$\%. 

The posterior is estimated using a Markov Chain Monte Carlo (MCMC) algorithm. Markov chains are random walks for which the transition probability from one state to the next state is independent of how the first state was populated. MCMC algorithms exploit the fundamental theorem of Markov Chains, which states that for long enough random walks, the length of time (which is equivalent to the probability) that a chain populates a specific state is independent of the initial state it started from. This set of limiting, long random walk, probabilities is called the stationary (or equilibrium) distribution of the Markov chain. Consequently, when a Markov chain is constructed with a stationary distribution equal to the posterior, $p(\theta | D)$, the samples drawn at every step during a sufficiently long random walk will closely approximate the posterior density. We adopt here the No U-Turn Sampler (NUTS), which is based on the Hamiltonian Monte Carlo (HMC) algorithm and is the default sampler used in the PyMC3 package \cite{python} in Python. For details about this sampler, see Hoffman and Gelman \cite{Hoffman:2014}.

We will estimate the number of signal counts as follows. The observed total counts, $T$, and our estimate for the observed background counts, $B$, summed over all channels of the peak region, are given by Equations~(\ref{eq:total}) and (\ref{eq:bkg}), respectively. We will write the Poissonian likelihoods for the signal and background as
\begin{equation}
\label{eq:bayes4}
p(T | s, b) = \frac{e^{-(s + b)} (s + b)^{T}}{T!} 
\end{equation}
\begin{equation}
\label{eq:bayes5}
p(B | b) = \frac{e^{-b} b^{B}}{B!} 
\end{equation}
Our model parameters are $s$ $\ge$ $0$ and $b$ $\ge$ $0$, which denote the true mean (expected) numbers of the signal and background counts, respectively. Equation~(\ref{eq:bayes4}) is justified on the grounds that the sum of two Poissonian random variables is again Poissonian distributed. The only assumptions made so far are that the signal and background are described by independent random variables and that the observed background is estimated from Equation~(\ref{eq:bg}). 

Priors need to be considered carefully in any Bayesian model. We expect that the choice of prior will matter little if the signal is large compared to the background. However, the prior will impact the results for low signal-to-noise ratios. In the following, we explore two very different priors and report the results for both choices. 

First, we chose a very broad, non-informative prior given by a half-Gaussian with its mode located at zero and a standard deviation of $\zeta$, i.e.,
\begin{equation}
\label{eq:bayes6a}
p(s) \propto e^{-s^2/(2\zeta_1^2)} 
\end{equation}
\begin{equation}
\label{eq:bayes7a}
p(b) \propto e^{-b^2/(2\zeta_2^2)} 
\end{equation}
Sensible choices for the standard deviations, $\zeta_j$, are values exceeding the total number of counts over the peak region (i.e., $\zeta_1$ $\approx$ $\zeta_2$ $\approx$ $1000$).

Second, previous work \cite{NARSKY2000444,narsky2000expected} has investigated different choices of priors for counting statistics in connection with upper bound estimates and found that the Jeffreys' prior provided a reliable description. For Poissonian likelihoods with mean values of $s$ and $b$, the Jeffreys' priors are given by
\begin{equation}
\label{eq:bayes6}
p(s) \propto \frac{1}{\sqrt{s}} 
\end{equation}
\begin{equation}
\label{eq:bayes7}
p(b) \propto \frac{1}{\sqrt{b}}
\end{equation}
The above functions are not normalizable (i.e., improper priors) and, therefore, do not represent probability densities. We will approximate them by using proper gamma probability distributions. For example, instead of Equation~(\ref{eq:bayes6}), we write
\begin{equation}
\label{eq:bayes8}
f(s, \alpha, \beta) = \frac{\beta^\alpha s^{\alpha - 1} e^{- \beta s}}{\Gamma(\alpha)} 
\end{equation}
with $\alpha$, $\beta$ $>$ $0$. The two parameters of the gamma distribution, $\alpha$ and $\beta$, are called ``shape'' and ``rate,'' respectively. The quantity $\Gamma(\alpha)$ denotes the gamma function,
\begin{equation}
\label{eq:bayes9}
\Gamma(\alpha) = \int_0^\infty x^{\alpha-1} e^{-x} dx
\end{equation}
The Jeffreys' priors of Equations~(\ref{eq:bayes6}) and (\ref{eq:bayes7}) can then be closely approximated by
\begin{equation}
\label{eq:bayes10}
p(s) \approx f(s, 0.5, 0.00001)
\end{equation}
\begin{equation}
\label{eq:bayes11}
p(b) \approx f(b, 0.5, 0.00001)
\end{equation}

In symbolic notation, our complete Bayesian model is given by
\begin{align}
&\text{Parameters:} \nonumber \\
&\quad \theta = (s, b) \nonumber \\
&\text{Likelihoods:} \nonumber \\
&\quad T \sim \mathrm{dpois}(s + b) \nonumber \\
& \quad B \sim \mathrm{dpois}(b) \nonumber \\
&\text{Prior choice 1:} \nonumber \\
&\quad s \sim \mathrm{dnorm}(0.0, \zeta_1^2)~T(0,) \nonumber \\
& \quad b \sim \mathrm{dnorm}(0.0, \zeta_2^2)~T(0,) \nonumber \\
&\text{Prior choice 2:} \nonumber \\
&\quad s \sim \mathrm{dgam}(0.5, 0.00001) \nonumber \\
& \quad b \sim \mathrm{dgam}(0.5, 0.00001) \nonumber \\
\end{align}
where ``dpois'', ``dnorm'', and ``dgam'' denote Poisson, normal, and gamma probability distributions, respectively. The notation ``T(0,)'' implies that only the right half of the Gaussian (i.e., positive values only) is used. The symbol ``$\sim$'' stands for ``has the distribution of.'' For the value of $s$, we will report the 50 percentile (median) of its posterior density, while the uncertainty of $s$ is estimated from the 16 and 84 percentiles (i.e., assuming a coverage probability of 68\%). 

If the number of signal counts from the above procedure is deemed statistically significant, we can estimate, in analogy to Method A, the peak centroid. The Bayesian model is the same as presented in Equations~(\ref{eq:bayes4}) - (\ref{eq:bayes11}), except that the formalism is now applied to each individual channel of the peak region. In other words, the model estimates the true signal counts, $s_i$, from the observed total counts, $T_i$, and background counts, $B_i$, for each channel, $i$. Similar to Equation~(\ref{eq:cent}), the centroid is then calculated at each step of the MCMC procedure by using
\begin{equation}
\label{eq:bayes12}
\bar{x} = \frac{\sum_i^n s_i x_i}{\sum_i^n s_i} 
\end{equation}
The value and uncertainty of $\bar{x}$ is again computed from the 16, 50, and 84 percentiles of its probability density.

Notice that it would be incorrect to estimate the total signal counts in the Bayesian model from $S$ $=$ $\sum_i s_i$, as implied by Equation~(\ref{eq:bayes12}). Suppose, zero counts are measured for both the total number of counts and the background. The Bayesian model presented above will yield a median value of $\approx$1.0 (using Prior 1) or $\approx$0.5 (using Prior 2) for the estimated signal counts. If, instead, we assume a peak region 20 channels wide, with zero total and background counts in each channel, summing the estimated signal counts would result in a nearly Gaussian-shaped probability density with a median value near $\approx$20 (for Prior 1) or $\approx$10 (for Prior 2). Clearly, this is a nonsensical result because the summed signal counts are biased towards a larger value. However, for the purposes of the centroid estimation only, this bias is significantly lessened because it affects both the numerator and denominator of Equation~(\ref{eq:bayes12}) in similar ways. 

If no obvious signal is present in the peak region, a proper Bayesian analysis will involve the comparison of two exclusive models: the null hypothesis that the observed counts are caused by background only versus the alternative hypothesis that a signal process contributes to the counts (see, e.g., Ref.~\cite{Knoetig_2014,Casadei:2015fd}, and references therein). As already mentioned in Section~\ref{sec:intro}, we are not concerned here with rigorous upper bound estimates, but our main focus are cases where a signal can be discerned in the data. To quantify the possibility that a signal is present, but is too weak to be measured, we will use the concept of the Highest (posterior) Density Interval (HDI). It summarizes the range of the most credible parameter values and is defined, for a given probability, by any parameter value inside the HDI having a higher probability density than any value outside the HDI \cite{box2011bayesian}. 

For the MCMC simulations using Method B, the variables of interest are sampled using a Markov chain of length 10,000 samples distributed over two chains, excluding a burn-in phase of $2000$ steps for each chain. This ensured convergence of all chains. 

{\it Comment 6:} For relatively small signal-to-noise ratios, the estimated number of signal counts will be impacted by the choice of prior.

\subsection{Method C: Gaussian fit} \label{sec:gaussfit}
The methods discussed so far have made no assumptions about the shape of the peak. If the peak shape is known, a suitable model can be fit to the number of counts in the peak and background regions to extract the estimates for the signal centroid and intensity.

In many cases, the detector response is accurately described by a Gaussian function
\begin{equation}
\label{eq:fit}
g_i(x_i) = a e^{-(x_i - b)^2 /(2c^2) } 
\end{equation}
where $x_i$ denotes the channel. The parameters $a$, $b$, and $c$ are the normalization (height), centroid, and standard deviation of the Gaussian, respectively. For our test data, we will assume that this Gaussian signal sits on top of a linear background, 
\begin{equation}
\label{eq:fit2}
f_i(x_i) = d x_i + e  
\end{equation}
which introduces two more parameters, $d$ and $e$, for the slope and intercept, respectively\footnote{For an application of a Bayesian method to determine the peak area in Ge detector spectra, see Ref.~\cite{LONGORIA1990308}. Their model describes physical processes, e.g., partial energy deposition, charge trapping, recombination, etc., and is especially useful for strong peaks on top of a small background. The usefulness of the Bayesian method for fitting a peak doublet measured in the decay of $^{101}$Mo has been demonstrated in Ref.~\cite{HAMMED1993543}.}. 

We will perform the fit using, again, a Bayesian model. The full model is now given in symbolic notations by
\begin{align}
&\text{Parameters:} \nonumber \\
&\quad \theta = (a, b, c, d, e) \nonumber \\
&\text{Likelihoods:} \nonumber \\
&\quad T_i \sim \mathrm{dpois}(g_i + f_i) \nonumber \\
& \quad B_i \sim \mathrm{dpois}(f_i) \nonumber \\
&\text{Priors:} \nonumber \\
&\quad a \sim \mathrm{dnorm}(0.0, \xi_1^2)~T(0,) \nonumber \\
&\quad b \sim \mathrm{dunif}(x_{p1}, x_{p2}) \nonumber \\
&\quad d \sim \mathrm{dnorm}(0.0, \xi_2^2) \nonumber \\
&\quad e \sim \mathrm{dnorm}(0.0, \xi_3^2) \nonumber \\
&\text{for strong peaks:} \nonumber \\
&\quad c \sim \mathrm{dnorm}(0.0, \xi_4^2)~T(0,) \nonumber \\
&\text{for weak peaks:} \nonumber \\
&\quad c \sim \mathrm{dnorm}(c^{exp}, [\sigma_c^{exp}]^2) \nonumber\\
\end{align}
where $T_i$ and $B_i$ are, same as before, the number of counts in each channel, $x_i$, of the peak and background regions, respectively; $g_i$ and $f_i$ are given by Equations~(\ref{eq:fit}) and (\ref{eq:fit2}), respectively; $\xi_1,...,\xi_4$ are chosen sufficiently large (e.g., $\xi_i$ $=$ $1000$) to describe the Gaussian height, standard deviation, and background by broad, non-informative priors. Unlike for Methods A and B, the background in Method C is, therefore, adjusted during the fit. The prior for $a$ is truncated at zero because the height of the Gaussian is a manifestly positive quantity. The uniform prior for the centroid, $b$, restricts the parameter search to the chosen peak region, $\left[ x_{p1}, x_{p2}\right]$. The quantities of main interest are the centroid, $b$, and area, $\sqrt{2\pi}ac$, of the Gaussian-shaped signal peak.

\begin{figure}
\includegraphics[width=0.85\columnwidth]{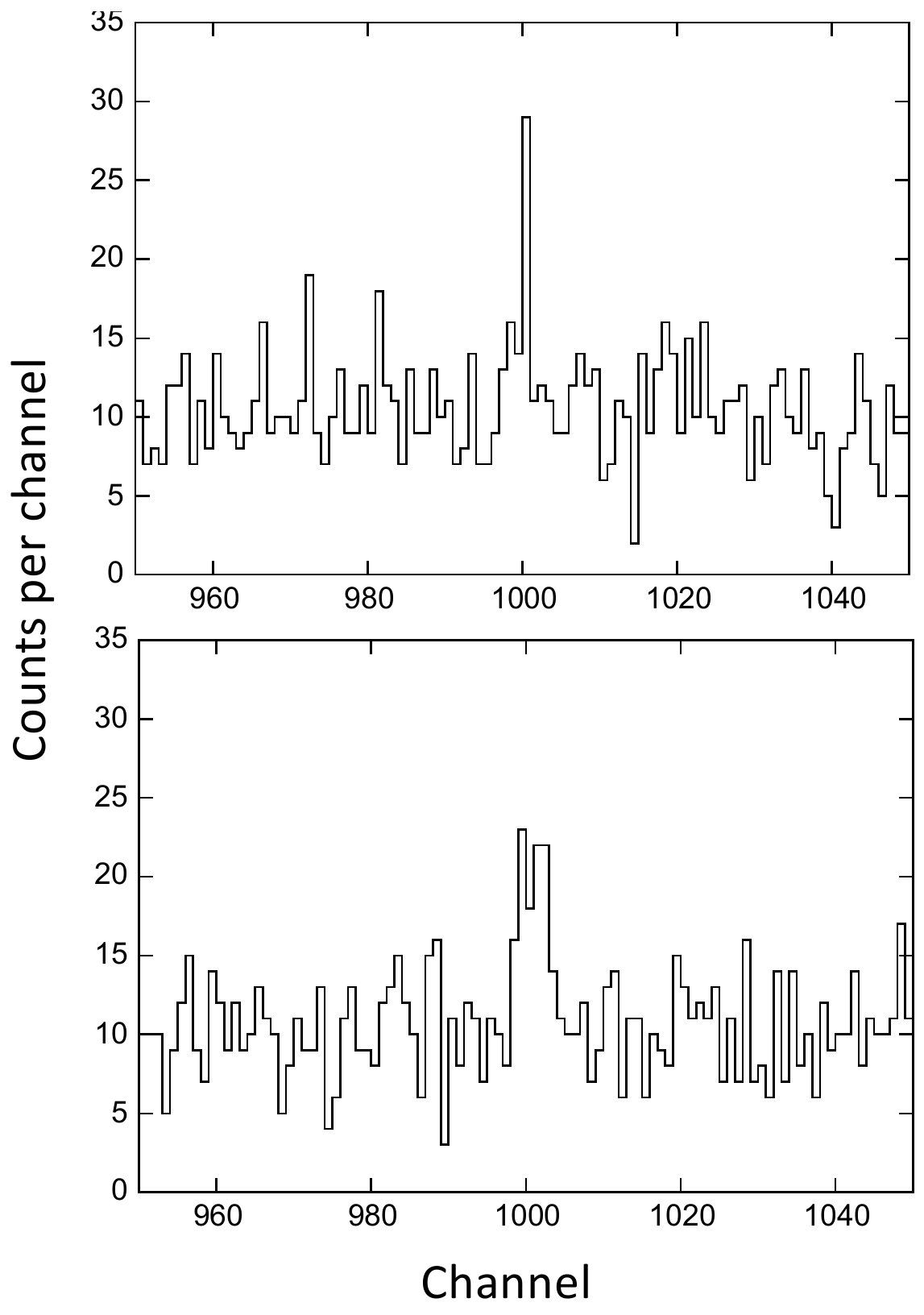}
\caption{\label{fig:datawidth} Two test spectra that were generated using exactly the same parameters (50 signal counts; centroid of channel 1000; background of 10 counts per channel; FWHM of 4.7 channels). The apparent width of the two peaks are markedly different, which is solely caused by statistical fluctuations.
}
\end{figure}

The model above has two options of priors for the Gaussian standard deviation, $c$, depending on the signal-to-noise ratio of the peak. For strong peaks, $c$ is an adjustable parameter used to estimate the detector resolution. As the signal strength decreases, or the background increases, the procedure of treating $c$ as an adjustable parameter becomes problematic. Figure~\ref{fig:datawidth} shows two test spectra that were generated with exactly the same values for the signal counts, centroid, background per channel, and FWHM. It can be seen that the apparent widths of the two peaks near channel $1000$ differ greatly, which is solely caused by statistical fluctuations. Clearly, fits with the standard deviation, $c$, as an adjustable parameter would result in markedly different widths of the fitted Gaussian function for the two spectra shown. Such a procedure introduces a bias, because the fitted peak widths may differ greatly from the detector response.

Therefore, we will first use for the analysis of the high signal-to-noise Data Set 1 a broad, non-informative prior to estimate the Gaussian standard deviation, $c^{exp}\pm\sigma_c^{exp}$. Subsequently, the lower signal-to-noise Data Sets 2 and 3 are analyzed using this result with an informative prior. 

Note, that the user-defined peak regions for Methods A and B had to be chosen as narrow as possible, while still containing the peak, so that the uncertainties on the signal counts are not inflated unnecessarily. On the other hand, Method C does not have that sensitivity because the peak shape is directly implemented into the model. 

The variables of interest are sampled using a Markov chain of length $50,000$, excluding a burn-in phase of $10,000$ steps. 

{\it Comment 7:} For peaks of low signal-to-noise ratios, treating the peak width as a freely adjustable parameter during the fit will introduce a bias. In such cases, it is more reasonable to estimate the Gaussian width by analyzing strong peaks and then to use the result in the subsequent analysis of weaker peaks.

{\it Comment 8:} Unlike the other methods, for a linear background the Gaussian fit method is insensitive to the choice of the width of the peak region, as long as this region is chosen sufficiently wide. 

\begin{table*}[]
\centering
\begin{threeparttable}
\caption{Analysis of test data.\tnote{a}}
\label{tab:results}
\begin{tabular}{l c c c c}
\toprule
  		&     \multicolumn{4}{c}{Method}  \\
  		&     A   &    B\tnote{b} (Prior 1)\tnote{c}    &  B\tnote{b} (Prior 2)\tnote{d}  & C\tnote{b}  \\ 
\midrule
  Test data 1    &         \multicolumn{4}{l}{True counts = 970; True centroid = 985} \\      
      &         \multicolumn{4}{l}{Actual counts = 948$\pm$33; Actual centroid = 984.91$\pm$0.06} \\      
Observed counts:	    	&   948$\pm$35	    &		947$\pm35$          &	948$\pm35$              & $950\pm33$\tnote{f} \\	
Observed centroid:	    &   984.89$\pm$0.06	&		984.86$\pm$0.08     &	984.87$\pm$0.08         & $984.86\pm0.07$\tnote{f} \\
\midrule
 Test data 2     &         \multicolumn{4}{l}{True counts = 105; True centroid = 985} \\      
      &         \multicolumn{4}{l}{Actual counts = 106$\pm$14; Actual centroid = 984.86$\pm0.19$ }\\      
Observed counts:	    	&   99$\pm$20	    &		99$\pm$20           &	97$\pm$20               & $97\pm14$\tnote{g} \\
Observed centroid:	    &   984.81$\pm$0.24	&		984.69$\pm$0.43     &	984.69$\pm0.48$         & $984.72\pm0.32$\tnote{g} \\
\midrule
 Test data 3     &         \multicolumn{4}{l}{True counts = 35; True centroid = 985} \\      
      &         \multicolumn{4}{l}{Actual counts = 38$\pm$10; Actual centroid = 985.00$\pm$0.32} \\      
Observed counts:	      	&   25$\pm$17   	&		27$\pm$16           & $\leq$52\tnote{e}  & $32\pm11$\tnote{g} \\
Observed centroid:	    &   985.31$\pm$0.29	&		984.92$\pm0.60$     &	                        & $985.16\pm0.77$\tnote{g} \\
\bottomrule
\end{tabular}
\begin{tablenotes}
\item[a]{All centroids are in units of channels. ``True'' values represent the starting parameters used to simulate the artificial data; ``actual'' values denote the signal counts and centroid calculated from Equations~(\ref{eq:netcounts}), (\ref{eq:netcountserr}), (\ref{eq:cent}), and (\ref{eq:centerr}) {\it before any background was added to the signal histogram}, i.e., for $B$ $=$ $0$; see Section~\ref{sec:test} for more information. The results listed are also shown in Figure~\ref{fig:summary}.}
\item[b]{Values and uncertainties are defined by the 16, 50, and 84 percentiles of the resulting posteriors.}
\item[c]{Results for a broad, half-Gaussian prior; see Equations~(\ref{eq:bayes6a}) and (\ref{eq:bayes7a}).}
\item[d]{Results for a Jeffreys prior; see Equations~(\ref{eq:bayes6}) and (\ref{eq:bayes7}).}
\item[e]{Upper bound defined by Equation~(\ref{eq:bayes3}), with $\theta_a$ $=$ $0$, $\theta_b$ $=$ $52$, and $\beta$ $=$ $97.5$\%.}
\item[f]{The Gaussian standard deviation, $c$, was estimated using a non-informative prior (see ``for strong peaks'' in Section~\ref{sec:gaussfit}), which resulted in $c^{exp}\pm\sigma_c^{exp}$ $=$ $1.95\pm0.06$ channels.}
\item[g]{The Gaussian standard deviation, $c$, was described by an informative prior (see ``for weak peaks'' in Section~\ref{sec:gaussfit}), using the result obtained under footnote \tnote{f}.}
\end{tablenotes}
\end{threeparttable}
\end{table*}

\begin{figure}[htbp]
\includegraphics[width=0.7\columnwidth]{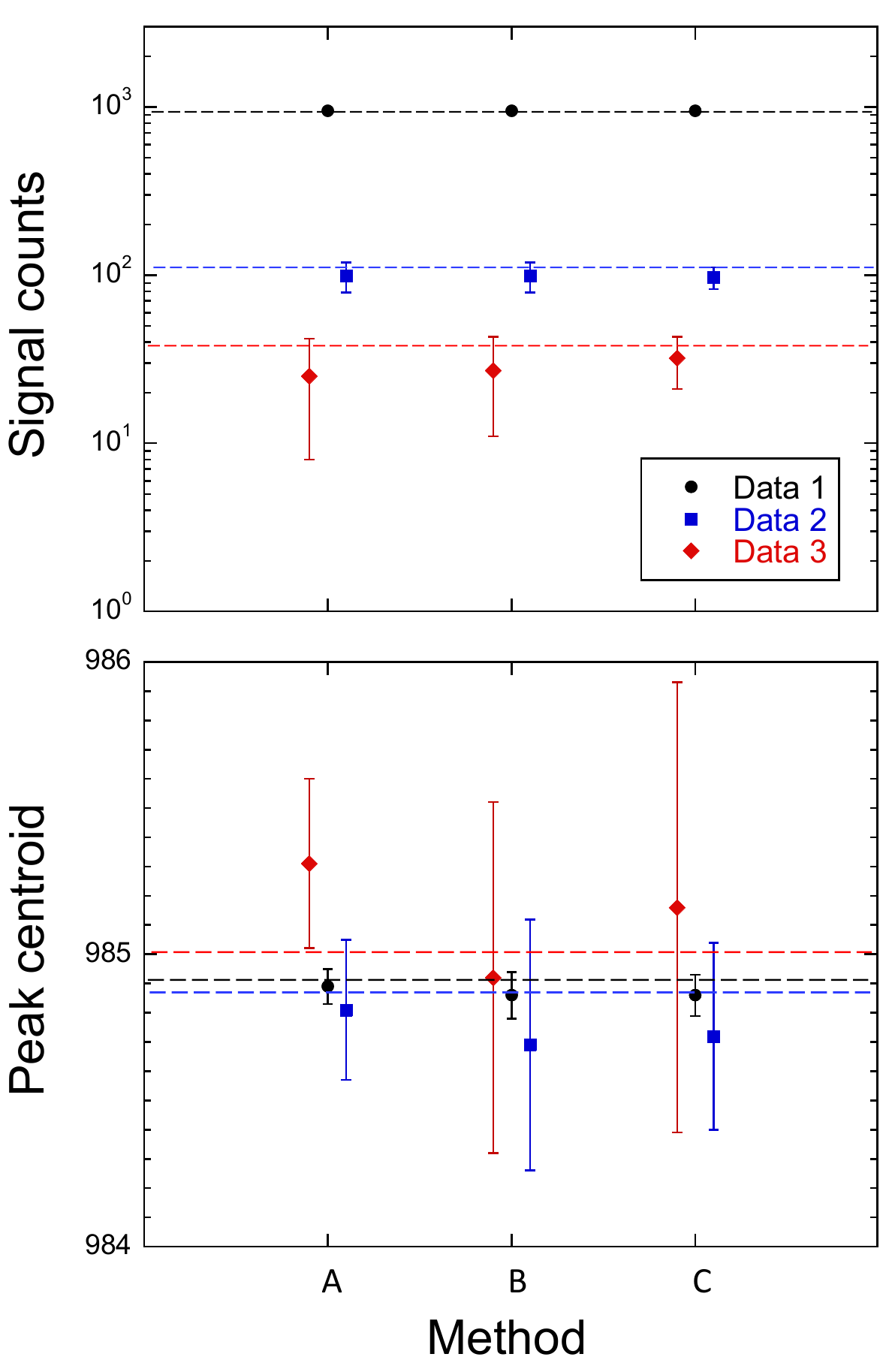}
\caption{\label{fig:summary} Visual representation of the results obtained from Methods A, B, and C (see Table~\ref{tab:results}). For Method B, only the values for Prior 1 are shown. (Top) Signal counts; (Bottom) Peak centroid. The black, blue, and red points depict the values for Data Sets 1, 2, and 3, respectively, and the color-coded horizontal dashed lines indicate the respective ``actual'' values (see Section~\ref{sec:test}). Interesting observations: (i) For low and moderate signal-to-noise data (red and blue points, respectively), Method C provides much smaller signal count uncertainties compared to the other two methods (top panel); (ii) For low and moderate signal-to-noise data (red and blue points, respectively), the more reliable Methods B and C provide larger peak centroid uncertainties than Method A (bottom panel).
}
\end{figure}


\section{Results}\label{sec:results}
We will discuss below the analysis of the artificial data (Section~\ref{sec:test}) using the methods outlined in Section~\ref{sec:method}. Usually, the detector resolution is known to the experimenter, which provides a natural choice for the width of the peak region to be analyzed. For Methods A and B, it makes little sense to choose a peak region width that exceeds the detector resolution by large factors since this would increase the uncertainty in the signal counts. As already mentioned above, Method C implements the line shape into the model and, therefore, is not sensitive to the width of the peak region. Here, we will use the same peak region for all three data sets and choose a region width of $14$~channels (channels $978$ $-$ $991$; see blue shaded regions in Figure~\ref{fig:data}), corresponding to three times the FWHM (Section~\ref{sec:test}). 

The background regions to the left and right of the peak region should be chosen wide enough to take statistical count fluctuations in the background into account. Here, we define channels $910$ $-$ $960$ and channels $1020$ $-$ $1070$ as left and right $50$-channel-wide background regions, respectively. We performed tests using a Monte Carlo method to assess if our results show any dependence on the chosen background regions (see Comment 5 in Section~\ref{sec:method}). To this end, we chose two 200-channel-wide regions on either side of the peak region. Within these wide regions, sub-regions were sampled randomly, each with a width of 50 channels. The background in the peak region was determined as before by using Equation~(\ref{eq:bg}) and the peaks are analyzed according to Method A. This process is then repeated many times. We found that our results were reproducible within uncertainties. In other words, our 50-channel-wide background regions were chosen sufficiently broad for taking statistical fluctuations into account.

Recall for the following discussion the distinction between ``true'', ``actual'', and ``observed'' values for the signal counts and the background (see Section~\ref{sec:test}). ``True'' values refer to the signal parameters used to generate the data. ``Actual values'' are determined from Equations~(\ref{eq:netcounts}), (\ref{eq:netcountserr}), (\ref{eq:cent}), and (\ref{eq:centerr}) after drawing signal counts from a Poissonian distribution for each channel of the Gaussian peak, but before any background is added (i.e., for $B$ $=$ $0$). For real data, neither the ``true'' nor the ``actual'' values are known. Our goal is to estimate ``observed'' values for the signal counts and the centroid using the different methods discussed in Section~\ref{sec:method}. 

Numerical results for Data Sets 1, 2, and 3 are listed in Table~\ref{tab:results} and are displayed in Figure~\ref{fig:summary}, where the horizontal dashed lines correspond to the actual values. 

\subsection{Analysis of Data Set 1}\label{sec:data1analysis1}
Data Set 1 has a high signal-to-noise ratio and was generated with $970$ true signal counts and a true centroid of channel $985.00$. The actual values  are 948$\pm$33 signal counts and a centroid at channel 984.91$\pm$0.06. 

{\it Method A:} We extract from the observed data 948$\pm$35 signal counts and a centroid of channel 984.89$\pm$0.06, in agreement with the actual values. 

{\it Method B:} We find 947$\pm$35 signal counts for both the half-Gaussian and the Jeffreys priors. The obtained centroids also agree for both priors. 

{\it Method C:} The Gaussian standard deviation is treated as an adjustable parameter, described by a non-informative prior. The resulting fit gives 950$\pm$33 signal counts, a centroid of channel 984.86$\pm$0.07, and a standard deviation of $c$ $=$ $1.95\pm0.06$ channels. The signal counts and centroid agree with the actual values.

{\it Summary:} For data of high signal-to-noise ratios, the results from all models are in agreement, as can be seen from the black points in Figure~\ref{fig:summary}. These results also agree with the actual values.

\subsection{Analysis of Data Set 2}\label{sec:data1analysis2}
Data Set 2 has a moderate signal-to-noise ratio. It was generated with $105$ true signal counts and a true centroid of channel $985.00$. The actual values are 106$\pm$14 signal counts and a centroid at channel 984.86$\pm$0.19. 

{\it Method A:} We find 99$\pm$20 signal counts and a centroid in channel 984.81$\pm$0.22, in agreement with the actual values. 

{\it Method B:} The half-Gaussian and Jeffreys priors result in 99$\pm$20 and 97$\pm$20 signal counts, respectively, and centroids in channels 984.69$\pm$0.43 and 984.69$\pm$0.48, respectively. In other words, both priors provide consistent results.  

{\it Method C:} An informative prior is used for the Gaussian standard deviation, using the value and uncertainty found in the analysis of Data Set 1. The derived signal counts and centroid amount to 97$\pm$14 and 984.72$\pm$0.32, respectively. These results agree with the actual values.

{\it Summary:} The results from Models A, B, and C are in agreement (see blue points in Figure~\ref{fig:summary}) and they are also consistent with the actual values. However, a number of interesting differences are apparent. 

First, Methods B and C provide centroid uncertainties significantly larger than those obtained from Method A (bottom panel of Figure~\ref{fig:summary}). In fact, the value from Method A ($\pm0.24$ channels) is only slightly larger than the actual centroid uncertainty ($\pm0.19$ channels). Recall that the latter represents the result of the signal {\it before} any background had been added. Since it is reasonable to assume that a significant background will increase the uncertainty of the peak centroid, it appears that simple count summation (Method A) underpredicts the centroid uncertainty. The underlying reason is that Equations~(\ref{eq:cent}), (\ref{eq:stand}), and (\ref{eq:centerr}) disregard the uncertainty in the signal counts, $S_i$, as already noted in {\it Comment 3} of Section~\ref{sec:method_A}. 

Second, Method C gives a signal count uncertainty much smaller than Methods A and B (top panel of Figure~\ref{fig:summary}). In fact, the value from Method C ($\pm14$) is the same as the actual signal count uncertainty. Therefore, it appears that the additional assumption of a given peak shape (here, a Gaussian) results in a signal count uncertainty similar to that obtained before the background was added to the signal histogram (i.e., the actual value).

\subsection{Analysis of Data Set 3}\label{sec:data1analysis3}
Data Set 3 contains $35$ true signal counts, with a true centroid of channel $985.00$. The actual values are 38$\pm$10 signal counts and a centroid at channel 985.00$\pm$0.32. 

{\it Method A:} We find 25$\pm$17 signal counts, which appears to agree with the actual value. Our result for the centroid, 985.31$\pm$0.29, exceeds the actual value by one standard deviation. See the red points in Figure~\ref{fig:summary}. 

\begin{figure}
\includegraphics[width=0.9\columnwidth]{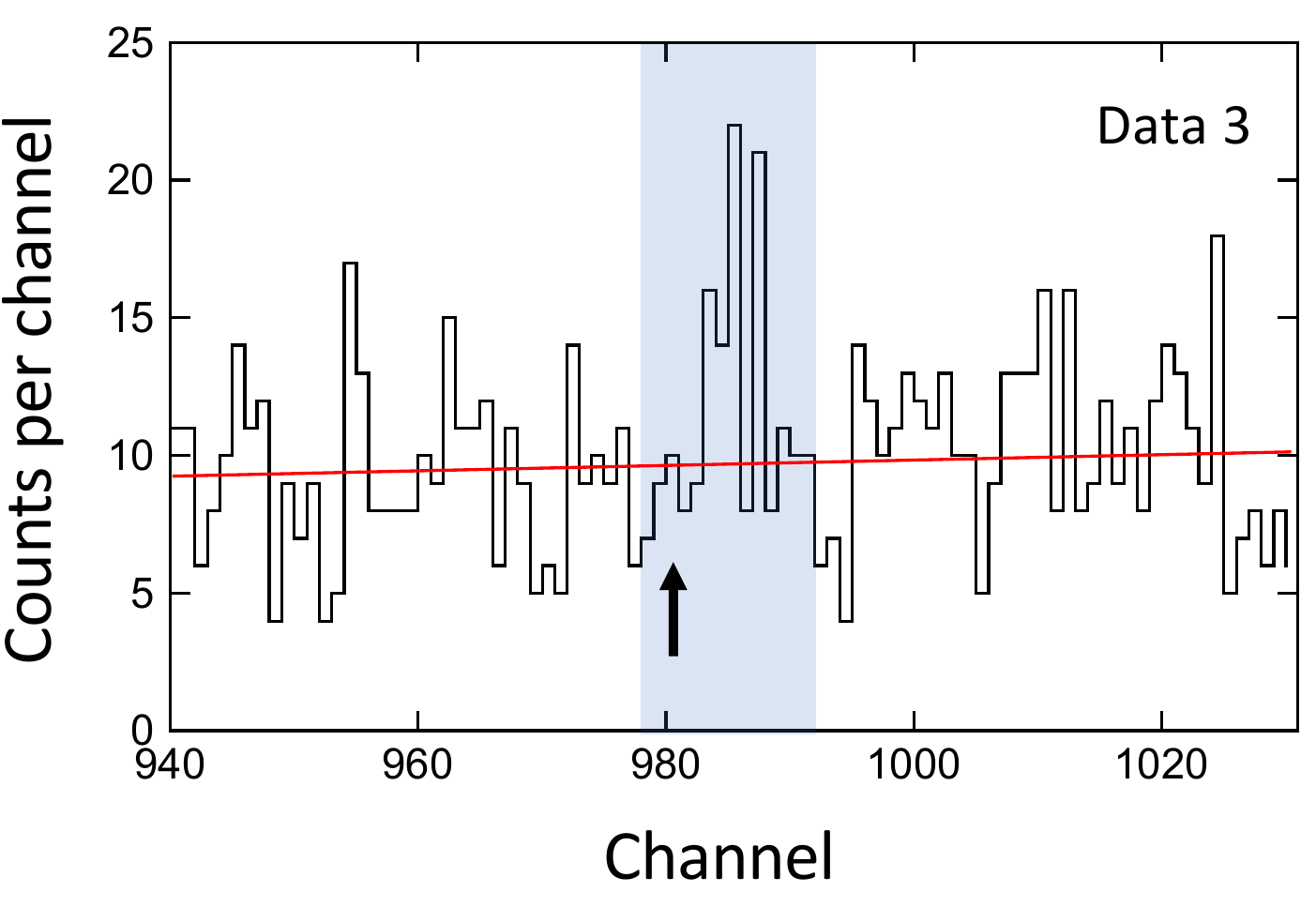}
\caption{\label{fig:data3} Expanded view of Data Set 3 (see bottom panel in Figure~\ref{fig:data}). The red line depicts the linear background estimated as discussed in Section~\ref{sec:method} and the peak region is shaded blue. Notice how the background is larger than the number of observed counts for several channels in the beginning of the peak region, near the arrow (see text).
}
\end{figure}

These results imply two problems for low signal-to-noise data, such as Data Set 3. The first problem is related to the bias introduced by channels in the peak region for which the total number of observed counts is smaller than the predicted background. This can be seen in Figure~\ref{fig:data3}, which depicts the data shown in the bottom panel of Figure~\ref{fig:data} on an expanded channel scale. The red straight line corresponds to the  background estimated according to Section~\ref{sec:method} and the peak region is shaded blue. A negative number of signal counts is obtained for four out of the first five channels in the peak region (near the black arrow). Including these negative signal counts in the sum of Equation~(\ref{eq:stand}) yields a negative value for $s_x^2$, which is nonsensical. Therefore, we disregarded those channels in the sum, as already pointed out in Section~\ref{sec:method_A} (see {\it Comment 4}). However, this procedure biases the uncertainty in the peak centroid, which appears to be much smaller compared to the other methods (see below).

The second problem is related to the comparable magnitudes for the mean value of the total signal counts and its associated uncertainty. A result such as ``25$\pm$17'' has no rigorous statistical meaning because Method A provides no information on the probability density associated with this outcome. Furthermore, it would be tempting to improve statistics by choosing a narrower peak region that includes only the channels with the highest number of counts. However, this procedure, which is without doubt frequently applied in practice, introduces yet another bias. As already mentioned, our peak region is 14 channels wide, equal to three times the (known) FWHM of the Gaussian used to generate the data. In practice, the FWHM corresponds to the detector resolution that is known to the experimenter. By choosing a peak region width that is significantly less than $3\times$FWHM, a large fraction of the peak area will be disregarded. Clearly, Method A has severe limitations when the signal-to-noise ratio is small because it provides no measure for the probability associated with the values obtained for either the signal counts or the centroid. 

{\it Method B:} With the half-Gaussian prior, we find 27$\pm16$ signal counts and a centroid of channel 984.92$\pm$0.60. The corresponding values for the Jeffreys prior are 17$^{+18}_{-14}$ signal counts and a centroid of channel 985.05$\pm$0.74. The values are derived from the 16, 50, and 84 percentiles of the posteriors (Section~\ref{sec:bayes}). 

\begin{figure}
\includegraphics[width=0.9\columnwidth]{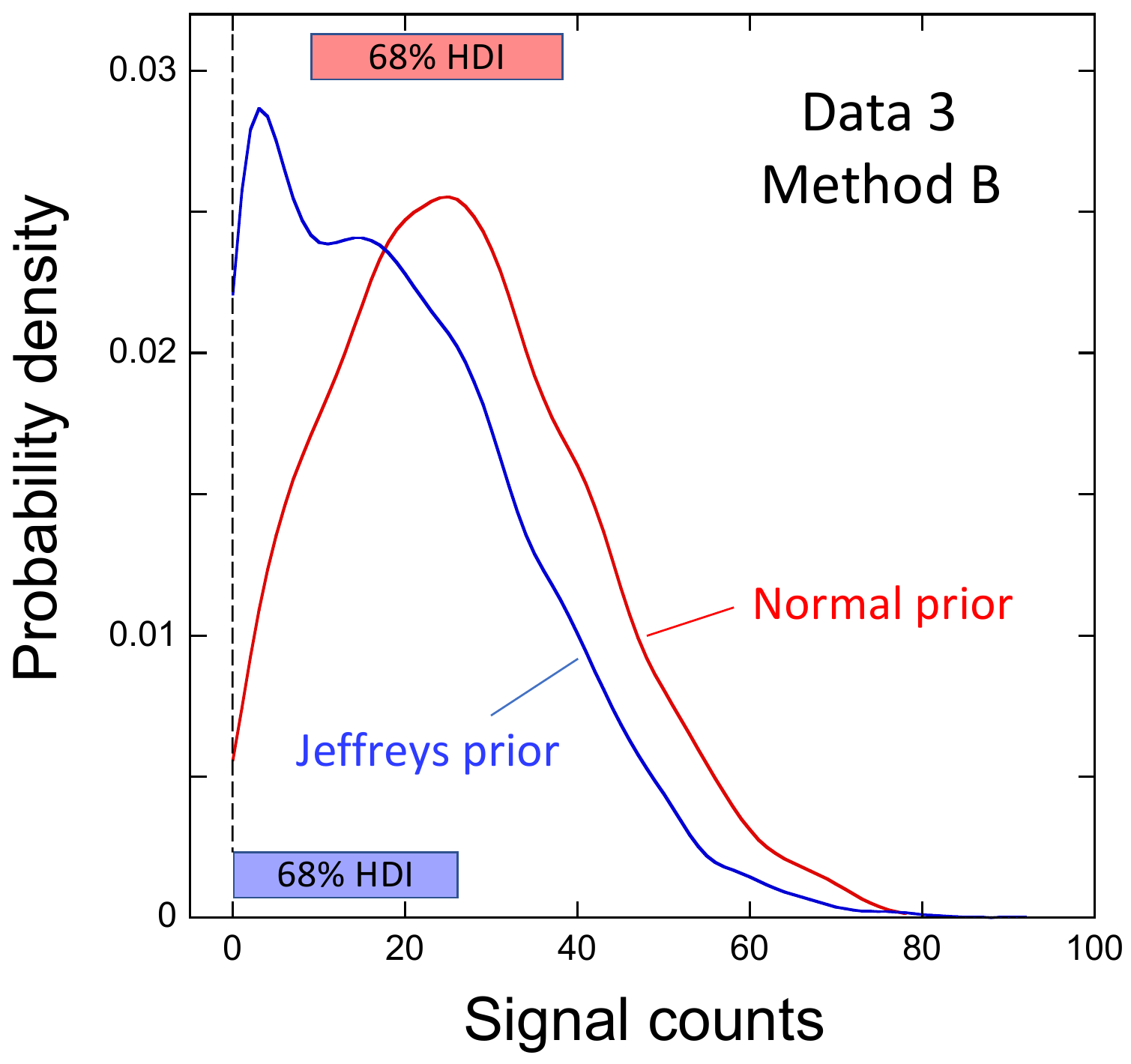}
\caption{\label{fig:bayes} Posteriors for the signal counts of the low signal-to-noise Data Set 3 (see Figure~\ref{fig:data3}) using Method B. The red and blue curves were obtained using a half-Gaussian (``normal'') and Jeffreys prior, respectively. The red (top) and blue (bottom) regions indicate 68\% HDI's (Highest Density Intervals) for the former and latter prior, respectively. Since the 68\% HDI for the Jeffrey's prior is consistent with zero signal counts, only an upper parameter bound should be presented.
}
\end{figure}

In contradistinction to the results from Method A, Model B allows for associating the derived values with probabilities. The posteriors for the two choices of priors are depicted in Figure~\ref{fig:bayes}. For the half-Gaussian prior, the posterior (red line) shows a resolved peak near $27$ signal counts and a significant decline in probability density towards zero signal counts. The red region (top) indicates the 68\% Highest Density Interval (HDI; see Section~\ref{sec:bayes}). Its lower bound is about 10 signal counts, indicating that a signal can be discerned and a parameter value, including uncertainties, can be presented. On the other hand, for the Jeffreys prior, the posterior (blue line) exhibits no resolved peak and a large probability density near zero signal counts. In this case, the lower bound of the 68\% HDI (blue region at bottom) includes zero signal counts. Therefore, it is more sensible in this case to present an upper bound instead of a central value with uncertainties. We find $\leq$52 signal counts with a coverage probability of 97.5\%, according to Equation~(\ref{eq:bayes3}).

As expected, the obtained values are sensitive to the choice of prior. Since we know that Data Set 3 was generated with $35$ signal counts, and we can also see visually a (albeit weak) peak in the spectrum (see bottom panel in Figure~\ref{fig:data}, and Figure~\ref{fig:data3}), we find that the choice of the half-Gaussian prior provides more reliable results. Using the Jeffreys prior, all that can reasonably be reported is an upper bound for the signal counts and thus no value for the centroid should be reported.

{\it Method C:} An informative prior is again used for the Gaussian standard deviation, using the value and uncertainty found in the analysis of Data Set 1. The derived signal counts and centroid amount to 32$\pm$11 and 985.16$\pm$0.77, respectively. These results agree with the actual values.

\begin{figure}
\includegraphics[width=0.9\columnwidth]{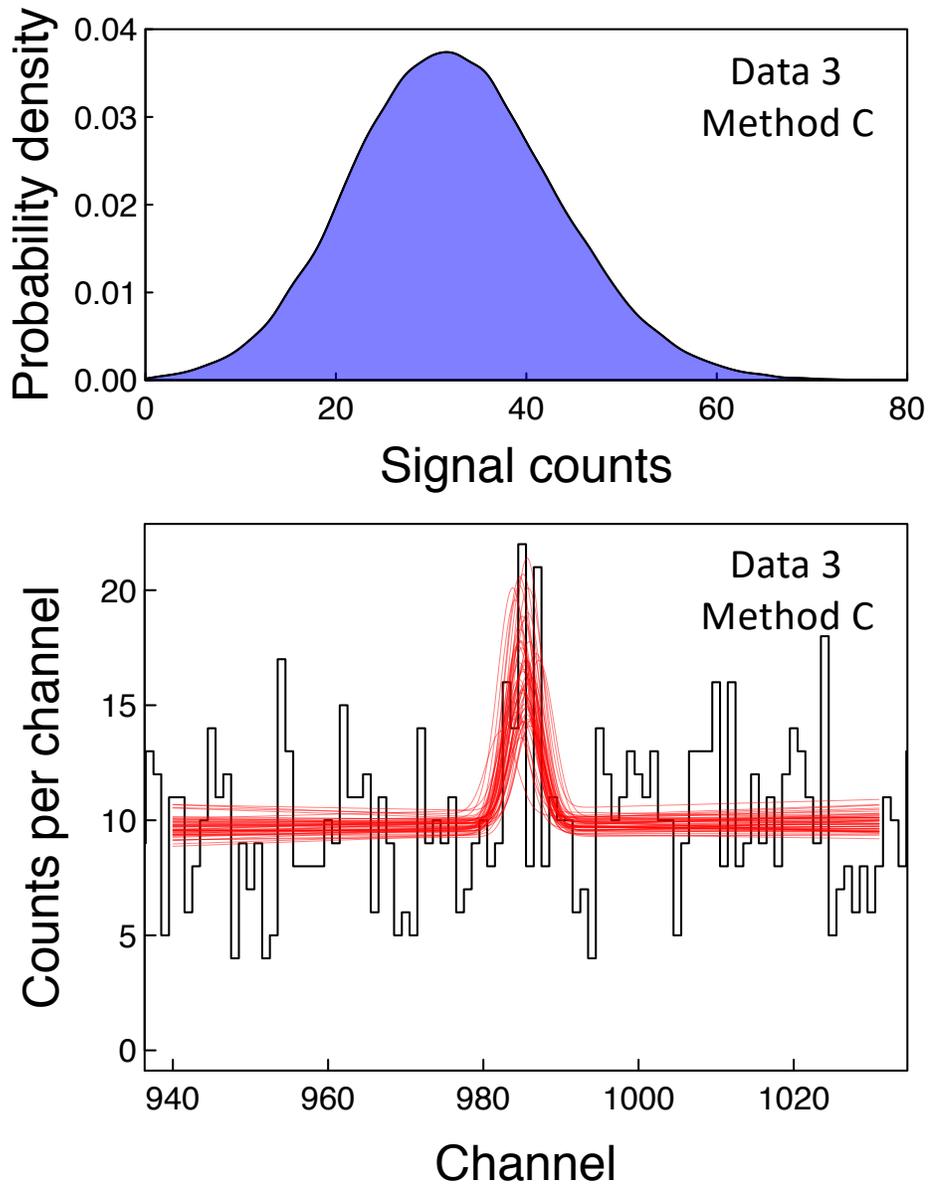}
\caption{\label{fig:gaussfit} Peak analysis of Data Set 3 (low signal-to-noise ratio) using Method C (Gaussian fit). (Bottom) 50 randomly chosen credible lines (from among $50,000$ samples). (Top) Probability density of the signal counts, as given by the areas under the Gaussian peaks shown in the bottom panel. Notice that the probability for zero signal counts is negligible.
}
\end{figure}

Results are shown in Figure~\ref{fig:gaussfit}. Part of the spectrum is depicted in the bottom panel, together with $50$ credible lines chosen randomly from among $50,000$ samples. The top panel displays the probability density of the signal counts, which is given by the areas under the Gaussian credible lines shown in the bottom panel. It can be seen that the probability density is single-peaked and the probability for zero counts is negligible.

{\it Summary:} For low signal-to-noise data, Method A, which is based on count summation, does not provide any probability estimates to assess if a signal is present in the data or not. With this method, a result such as ``25$\pm$17'' signal counts has no statistical meaning. Furthermore, Method A severely underpredicts the uncertainty in the peak centroid for the reasons outlined above.

Methods B and C provide probability densities that are used to define statistically meaningful values for the signal counts and the centroid. For Method B, adoption of the normal prior is preferred over the Jeffreys prior, as discussed above. As was the case for Data Set 2, the signal count uncertainty from Method C ($\pm11$) is similar to the actual signal count uncertainty ($\pm10$). Again, it appears that the additional assumption of a given peak shape (here, a Gaussian) results in a signal count uncertainty similar to that obtained before the background was added to the signal histogram (i.e., the actual value).

\subsection{Trends with additional data sets}\label{sec:trends}
\begin{figure}
\includegraphics[width=0.9\columnwidth]{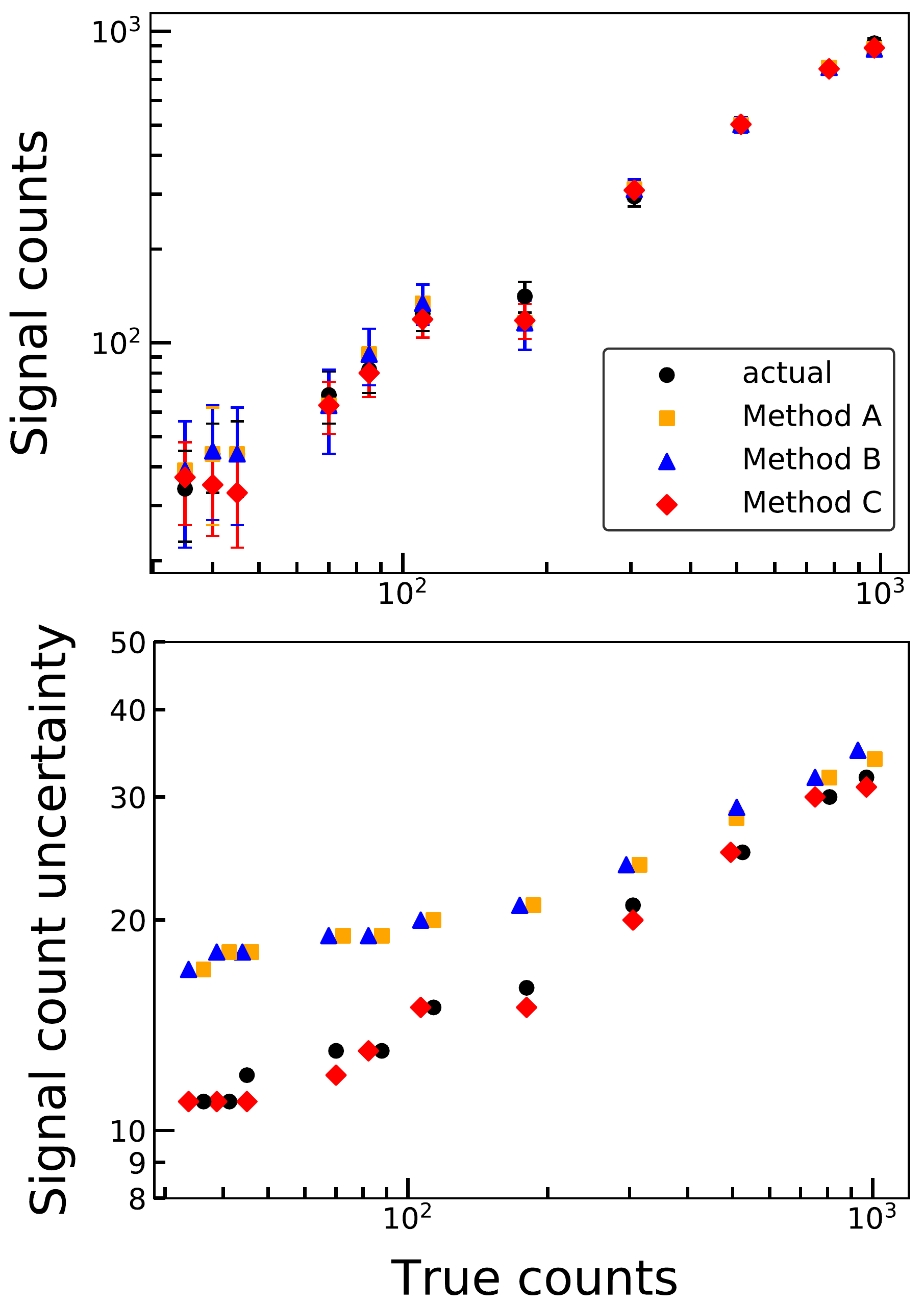}
\caption{Signal counts (top panel) and signal count uncertainty (bottom panel) versus true signal counts used to generate the artificial data. Black, yellow, blue, and red symbols correspond to results for the actual values, Method A, B, and C, respectively. 
}
\label{fig:summarySIG}
\end{figure}

\begin{figure}
\includegraphics[width=0.9\columnwidth]{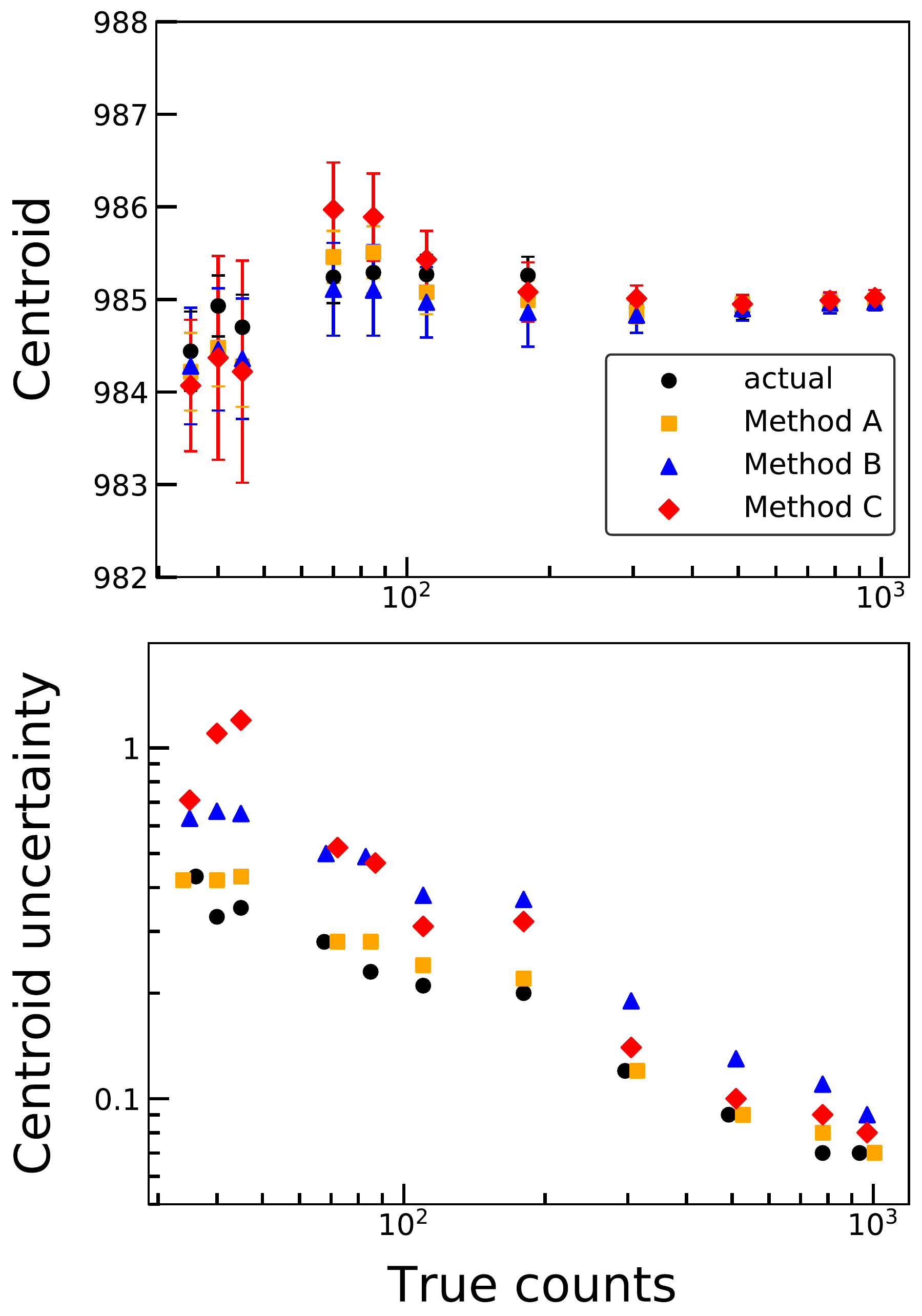}
\caption{Centroid (top panel) and centroid uncertainty (bottom panel) versus true signal counts used to generate the artificial data. The y-axis is in units of channels. Black, yellow, blue, and red symbols correspond to results for the actual values, Method A, B, and C, respectively. The true peak centroid is channel $985.0$.
}
\label{fig:summaryCENT}
\end{figure}

To check on our results discussed above and to reveal trends for data generated with a varying number of true signal counts, we show in the top panels of Figures~\ref{fig:summarySIG} and \ref{fig:summaryCENT} the actual (black) signal counts and centroids, respectively, and the observed values obtained with Method A (yellow), B (blue), and C (red). The bottom panels in the two figures show the uncertainties and make the trends more apparent.

For the observed signal counts (Figure~\ref{fig:summarySIG}), it can be seen that the uncertainties from Methods A and B are very similar in magnitude, independent of the true signal counts. The uncertainties from Method C, which are significantly smaller than those from Methods A and B, agree with the actual uncertainties over the entire range shown. We conclude that Method C provides superior results for signal counts compared to other methods.

For the observed centroids (Figure~\ref{fig:summaryCENT}), the uncertainties of Methods B and C are similar in magnitude, and larger than those from Method A or the actual values. The fact that Method A provides uncertainties in agreement with the actual values is most likely caused by the bias of disregarding channels with a negative number of signal counts. We conclude that either Method B or C will provide reliable results for the centroid uncertainty. 

\section{Summary and conclusions}\label{sec:summary}
We discussed the estimation of signal centroid and signal counts in the presence of background counts using three different methods. 

Method A is based on count summation. The analysis is straightforward, and does not involve any statistical modeling. It provides reliable results only for high signal-to-noise data. In all other cases, it has severe limitations. First, it gives significantly larger signal count uncertainties compared to Method C. Second, it provided centroid uncertainties that are biased towards values that are too small. Third, for low signal-to-noise data, the results are not quantifiable in terms of probabilities. For example, a result such as ``25$\pm$17'' signal counts has no rigorous statistical meaning. 

Method B employs a Bayesian model to extract signal counts and centroid from the measured total and background counts. The resulting values are derived from the respective posteriors and, therefore, have a rigorous statistical meaning that are associated with user-defined probabilities. The method makes no assumptions about the peak shape. As expected, for low signal-to-noise data the results are sensitive to the choice of priors for the signal and background counts. We found that assuming a non-informative prior based on a half-normal probability density reproduces the original values used to generate the artificial data. Method B yields reliable and relatively small centroid uncertainties. A limitation shared with Method A is that it provides significantly larger signal count uncertainties compared to Method C. 

Method C makes a strong assumption regarding the peak shape by fitting a Gaussian function to the data. The fit is, again, based on a Bayesian model. Unlike for Methods A and B, the method is insensitive to the choice of the width of the user-defined peak region. However, it does require careful consideration of the Gaussian width used in the fitting, which is usually given by the detector resolution. For high signal-to-noise data, the prior for the width should be non-informative, so that the Gaussian width  can be found directly from the fit to the data. For low or moderate signal-to-noise data, the prior of the Gaussian width should be informative and based on the result obtained in the analysis of the high signal-to-noise data. Under these conditions, Method C provides reliable and relatively small uncertainties both for the signal counts and the centroid.

We would like to thank Emily Churchman, Gaither Frye, Richard Longland, Caleb Marshall, and Federico Portillo for helpful comments. This work was supported in part by the U.S. DOE under contracts DE-FG02-97ER41041 (UNC) and DE-FG02-97ER41033 (TUNL).



\bibliographystyle{elsarticle-num} 
\bibliography{paper.bib}





\end{document}